# Optical Magnetic-Field Gating of Photon Avalanching Quantum Processes

Benoît Reynier[1°], Filippo Calavaro[1°], Natalie Fardian-Melamed[2], Eric Charron[1], Bleuenn Belz[1], Maria Sanz-Paz[1], Artiom Skripka[3,4], Sébastien Bidault[5], Emory M. Chan[3], P. James Schuck[2] and Mathieu Mivelle[1*]

[1] Sorbonne Université, Centre National de la Recherche Scientifique, Institut des NanoSciences de Paris, 75005 Paris, France

[2] Department of Mechanical Engineering, Columbia University, New York, New York 10027, USA

[3] The Molecular Foundry, Lawrence Berkeley National Laboratory, Berkeley, California 94720, USA

[4] Department of Chemistry, Oregon State University, Corvallis, Oregon 97331, USA

[5] Institut Langevin, ESPCI Paris, Université Paris Sciences et Lettres, Centre National de la Recherche Scientifique, 75005 Paris, France

° Equal contribution

[*]Corresponding author (mathieu.mivelle@sorbonne-universite.fr)

**Abstract**

Highly nonlinear phenomena, where macroscopic changes are triggered by microscopic perturbations, are ubiquitous in complex systems ranging from epidemic spreading to neural networks. In optics, photon avalanching represents a prime manifestation of such highly nonlinear behavior, characterized by a steep nonlinear response (s) following an avalanche threshold. However, this nonlinearity is typically determined by material composition, making this behavior challenging to tune for adaptive sensing, imaging, and neuromorphic applications. Here, we demonstrate giant optical magnetic-field-gated threshold and nonlinearity in lanthanide-doped nanoparticles. We exploit the distinct quantum selection rules of the avalanche process: while the energy-looping cycles of excited-state absorption and cross-relaxation are primarily electrically driven, the seed excitation—ground-state absorption—is a mixed transition permitting magnetic dipole interaction. By positioning the emitters within an optical standing wave to spatially partition the electromagnetic field, we tune the local magnetic-to-electric field balance to selectively regulate the seeding and amplification probabilities. This allows us to demonstrate an unprecedented dynamic range of control over the power threshold and nonlinearity, tuning the order from s = 15 to s = 62. Furthermore, we harness this spatial separation of electric and magnetic fields using tightly focused cylindrical vector beams to locally sculpt the nonlinearity of a single nanocluster, achieving sub-wavelength super-resolution imaging that substantially surpasses conventional photon-avalanche imaging under Gaussian illumination. These findings establish the magnetic component of light as a practical knob to actively program nonlinear optical dynamics, sub-diffraction nanoscopy, and critical quantum dynamics at the nanoscale.

## Introduction

Nature and society are governed by complex collective behaviors that operate near a tipping point.[1, 2, 3] Whether it is the cascading failure of financial markets,[4] the viral spreading of information on social media,[5] or the synchronization of neural firing,[6, 7] these systems share a defining feature: a highly nonlinear threshold. A small fluctuation in the input can trigger a disproportionately large, macroscopic transition. The 'steepness' of this response—quantifying how abruptly the system switches states—is one of the key metrics determining the system's sensitivity and stability. In condensed matter physics, photon avalanching (PA) represents a particularly salient optical manifestation of extreme nonlinearity..[8, 9, 10, 11] Through a chain reaction between ions, PA converts linear excitation into highly nonlinear emission,[12, 13, 14, 15, 16] offering a steepness (nonlinearity order s) far exceeding conventional multi-photon processes.[17, 18, 19, 20, 21, 22, 23]

However, whereas biological or socio-economic systems allow external parameters to directly shift a specific microscopic interaction rate (e.g., changing the transmission rate of a virus),[24] the intrinsic nonlinearity of PA is largely governed by the crystal lattice and doping concentration fixed during synthesis.[25] Harnessing the extreme sensitivity of this avalanche response has recently enabled extraordinary sensing capabilities, such as optical nanosensors capable of measuring piconewton forces,[26] while external strategies—including thermal, spectral, temporal, or plasmonic modulation—have been developed to tune the PA emission.[16, 27, 28, 29, 30] Yet, there has been no mechanism to actively control the degree of avalanche dynamics (nonlinearity and threshold) by selectively gating the individual microscopic transitions through the fundamental components of the optical field.

To achieve this control, we turn to a fundamental component of light that is typically disregarded: the optical magnetic field (H). While modern photonics largely utilizes electric dipole interactions—relying on the electric field (E)—rather than magnetic dipole interactions, we demonstrate that the optical magnetic field can effectively regulate avalanche responses. We exploit a specific quantum mechanical asymmetry derived from the lanthanide's electronic structure. The key mechanism sustaining PA consists of a feedback loop of excited-state absorption (ESA) and cross-relaxation (CR) energy transfer among excited lanthanide ions.[13] In $Tm^{3+}$-doped avalanching nanoparticles (ANPs), this “energy looping” cycle is governed primarily by electric dipole transitions, whereas the non-resonant ground-state absorption (GSA) that initiates PA is permitted by magnetic dipole selection rules. We hypothesized that we could leverage these differences in the quantum character of initiation and energy looping in PA to tune its dynamic behavior.

Here we demonstrate that by positioning the nanoparticles within an optical standing wave, we can spatially split the electromagnetic field components. This allows us to use the local magnetic-to-electric field balance to selectively regulate the seed excitation, effectively gating the onset of the avalanche. Consequently, we achieve a giant, continuous modulation of both the avalanche threshold and its steepness, varying the nonlinearity order s from 15 to 62. Crucially, unlike conventional active optical control—which typically relies on short pulses to transiently alter carrier densities in a material—our approach delicately manipulates the separate electric and magnetic field components of the excitation wave. To showcase the practical utility of this mechanism, we further implement spatial electric-and-magnetic field partitioning in far-field scanning microscopy using radially and azimuthally polarized cylindrical vector beams. By isolating a pure longitudinal electric field at the focal center under radial polarization, we locally boost the degree of nonlinearity experienced by a single nanocluster, demonstrating sub-wavelength super-resolution down to 180 nm—far exceeding the resolution achieved through standard Gaussian-beam photon avalanching in similar conditions. This tunability establishes a general framework where the optical magnetic field governs the dynamics of collective systems. This capability opens up exciting prospects for applications ranging from adaptive neuromorphic computing to critical sensing.[26, 31, 32]

To establish the physical basis of our approach, we first detail the photon avalanche mechanism in trivalent thulium ions ($Tm^{3+}$) under 1064 nm excitation.[13] As illustrated in Figure 1A, the process relies on an initial seed excitation: the Ground State Absorption (GSA). Reaching a critical population of these excited states subsequently triggers Excited State Absorption (ESA) followed by a cross-relaxation (CR) energy transfer with a neighboring ground-state ion. This interaction is the amplification step: it transfers energy from one highly excited ion to two ions initially in their ground state, effectively doubling the population of an intermediate excited state. These ions then participate in further ESA and cross-relaxation cycles. After repeating this feedback loop over numerous cycles, this photophysical chain reaction leads to an exponential accumulation of energy within the sensitizer network—the avalanche—culminating in a highly nonlinear enhancement of photon emission. The pivotal feature of this system lies in the distinct quantum selection rules of the transitions involved (Figure 1A, inset). The GSA ignition step ($^3H_6$ to $^3H_5$) is a mixed transition mediated simultaneously by electric (d) and magnetic (μ) dipoles,[33, 34] rendering it sensitive to both components of the electromagnetic field.[34] Conversely, the resonant ESA transition ($^3F_4$ to $^3F_2$) that fuels the avalanche cycle is forbidden by magnetic dipole selection rules, and thus governed exclusively by electric dipole interactions.

To manipulate the magnetic field in a monolayer of ANPs, we constructed an experimental setup (Figure 1B) that spatially partitions the electromagnetic field components at the nanoscale—a significant technical challenge requiring precise near-field optomechanical control to avoid the spatial averaging typical of macroscopic measurements.[35] A gold mirror nanofabricated at the apex of a near-field optical probe (upper inset) is suspended above a dense monolayer of ANPs supported on a glass substrate (lower inset). A linearly polarized 1064 nm beam excites the system from below the ANPs, with the electric and magnetic field vectors represented by orthogonal arrows. When reflecting off the gold mirror, the incident light forms a standing wave interference pattern, visualized schematically by the spatial field distribution lobes.[36] Crucially, the distance Z between the mirror and the nanoparticle layer is actively controlled via the microscope's piezoelectric stage (indicated by the vertical arrow). By modulating this gap, we shift the standing wave pattern relative to the fixed monolayer, thereby allowing us to selectively position the nanoparticles in a regime of maximum electric field, maximum magnetic field, or a mixed electromagnetic environment.[36] Finally, the resulting nonlinear emission—the photon avalanche—is depicted by the photons going downwards from the excited layer (see detailed experimental setup and the corresponding spectrum in SI).

Because the seed absorption (GSA) and the amplification loop (ESA + CR) obey different electric and magnetic dipole selection rules (see detailed microscopic energy-level diagrams in Supplementary Information), selectively partitioning the local electric and magnetic fields is expected to alter the balance between initiation and energy looping. Figure 1C illustrates the anticipated impact of this electromagnetic mode selection on the macroscopic avalanche response (Luminescence vs. Incident Power Flux). By positioning the emitters at electric-field antinodes (blue), magnetic-field antinodes (green), or in a far-field reference regime (red), we expect that manipulating the relative rates of electric and magnetic dipole transitions will substantially modify the avalanching behavior—shifting both the power threshold ($\Phi_E$, $\Phi_0$, $\Phi_H$) and the nonlinear steepness ($s_E$, $s_0$, $s_H$) of the transition.

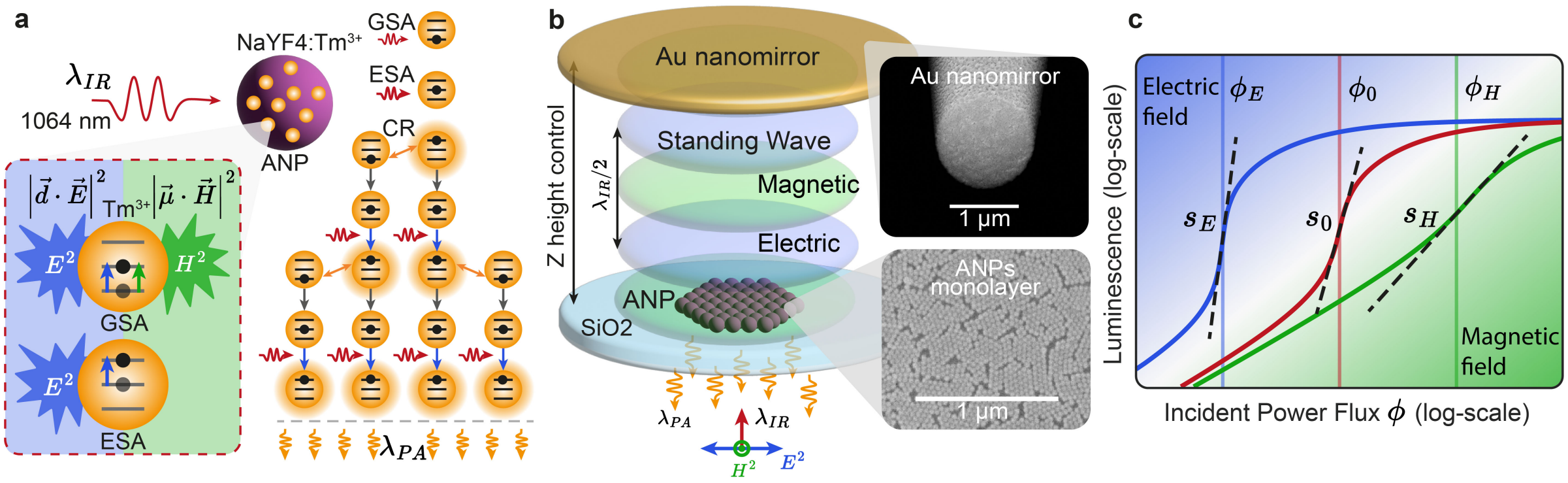

**Fig. 1.** ***Principle of optical magnetic-field-gated photon avalanching.*** *a) Mechanism of the photon avalanche in $Tm^{3+}$ ions. The process relies on a rare Ground State Absorption (GSA) 'ignition' followed by cooperative Excited State Absorption (ESA) (blue arrows) and cross-relaxation (CR) energy transfer cycles. Inset: Quantum selection rules. The GSA is a mixed transition permissible by both electric (d) and magnetic (μ) dipoles, whereas the ESA amplification loop component is strictly electric-dipole driven. b) Nanophotonic standing-wave setup. A gold mirror nanofabricated at the apex of a near-field probe (SEM inset) is suspended above a monolayer of avalanching nanoparticles. A 1064 nm excitation beam forms a standing wave where the electric (E) and magnetic (H) fields are spatially partitioned. Piezoelectric control of the mirror height (Z) selectively positions the monolayer in an E-field or H-field antinode. c) Conceptual macroscopic framework. Schematic luminescence power curves (log–log scale) illustrating how selectively driving electric-dipole (blue, $\Phi_E$, $s_E$) versus magnetic-dipole (green, $\Phi_H$, $s_H$) transitions relative to the far-field regime (red,*

$\Phi_0$, $s_0$*) is expected to modulate both the avalanche threshold (Φ) and the nonlinearity order (s).*

To rigorously quantify the impact of the field topology on the avalanche dynamics, we developed a theoretical model combining optical analytical calculations with a full set of population rate equations.[37, 38, 39] This approach allows us to calculate the corrected excitation rates of both GSA and ESA, due to the local electric ($|E|^2$) and magnetic ($|H|^2$) field intensities experienced by the nanoparticles and feed these values into the differential equations governing the $Tm^{3+}$ energy level kinetics (See Supplementary Information).

Figure 2A presents the resulting theoretical photon avalanche response curves, plotted on a double-logarithmic scale. A striking evolution is observed as the excitation mode shifts. The electric-dominated curves (blue) exhibit the signature of a highly efficient avalanche: a low power threshold and a steep, nearly vertical rise. However, as the magnetic field dominates, the curves undergo a twofold transformation. First, the avalanche threshold shifts towards significantly higher intensities, reflecting the difficulty of igniting the process using purely MD GSA. Second, the slope of the transition flattens dramatically. This theoretical prediction confirms that magnetic gating alters the avalanche dynamics.

This progressive reduction of cooperative feedback is quantified in Figure 2B, which plots the derivative (nonlinearity order s) of the population curves. The highest avalanche dynamics (stronger nonlinearity and lower threshold) are achieved under pure electric excitation, while magnetic coupling systematically dampens the response.

To unambiguously disentangle the specific roles of field intensity versus field ratio, we analyze the threshold and slope separately in Figures 2C and 2D. Figure 2C plots the calculated avalanche threshold $\Phi_A$ as a function of the mirror position (black dots). The data strictly follows the inverse of the local electric field distribution (blue curve), highlighting that the energetic onset of the avalanche is dictated only by the local electric field intensity. Notably, the standing-wave equipartition point (black star) exhibits a lower threshold than the far-field reference (red line). This difference is purely an intensity effect (here by a factor of 2), driven by the constructive interference boost from the mirror which enhances the local field density.

Conversely, Figure 2D displays the maximum nonlinearity slope *s* as a function of the electric-to-magnetic field mixture. Here, a crucial invariance appears: the far-field reference (red line) intersects the standing-wave data exactly at the equipartition point (black star). Despite their difference in threshold observed in Fig. 2C, both regimes yield identical nonlinearity because they share the same unity electric-to-magnetic field ratio ($\epsilon_0|E_0|^2 = \mu_0|H_0|^2$).

This confirms a fundamental law of the system: while the absolute electric field intensity dictates the avalanche threshold (governing the energetic onset and required pump power to ignite the cooperative process), the electric-to-magnetic field ratio governs the nonlinearity(determining the amplification efficiency and the steepness of the nonlinear response) (See more details in the Supplemental Information). By preserving the nonlinearity factor (s) correlation despite the threshold shift, the model demonstrates that the giant tuning observed is indeed driven by spatially exploiting the intrinsic quantum selection rules of the system, and not merely by photonic field enhancement.

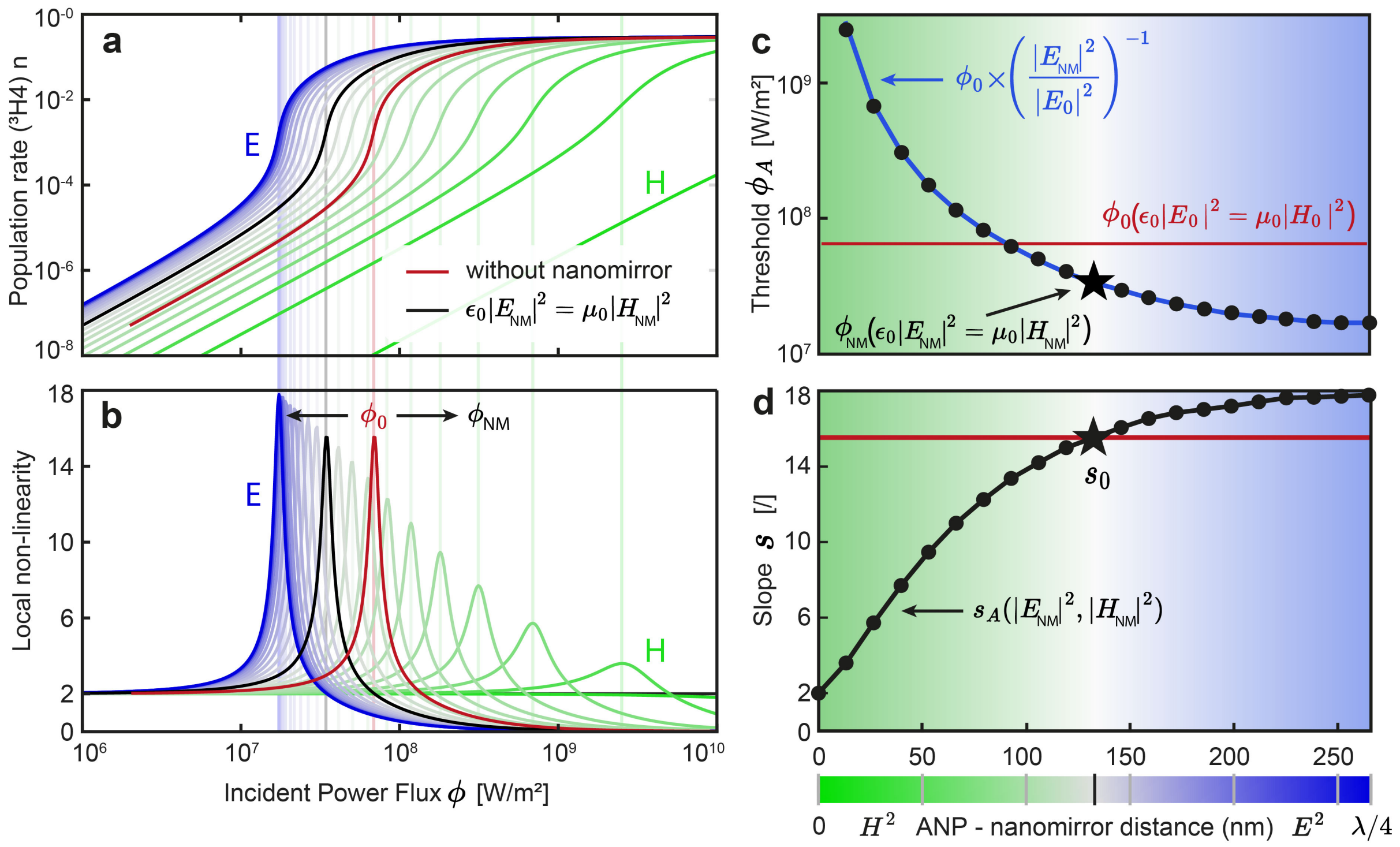

**Fig. 2.** ***Theoretical modeling of electromagnetic-gated avalanche dynamics.*** *a) Simulated photon avalanche response curves (log-log scale) calculated using a theoretical and rate-equation model. The y-axis shows the steady-state population of the 800 nm emitting level ($^3H_4$). The color gradient from blue (electric-dominant) to green (magnetic-dominant) reveals a progressive suppression of the avalanche steepness and a translation of the threshold to higher intensities. b) Nonlinearity order analysis. The derivative of the population curves reveals two concurrent effects: a*

*significant shift in the avalanche threshold (lower at electric anti-nodes, higher at magnetic anti-nodes) and a collapse of the nonlinearity factor (s) as the magnetic contribution increases. Furthermore, the comparison between the far-field regime (red curve) and the standing-wave equipartition point (black curve) demonstrates that while total field intensity affects the threshold position, the maximum nonlinearity is strictly governed by the local* $|E_{NM}|/|H_{NM}|$ *ratio. c) Calculated avalanche threshold in presence of the nanomirror* $\Phi_{NM}$ *plotted against the mirror position (black circles). The threshold strictly follows the inverse of the local electric field distribution (blue curve), highlighting that the PA ignition point depends on the local electric field intensity alone. The horizontal red line represents the far-field reference, while the black star marks the specific mirror position yielding equal electric and magnetic field amplitudes. (D) Corresponding maximum slope plotted as a function of the electric-to-magnetic field mixture. Similarly, the red line indicates the far-field value, and the black star denotes the equipartition condition.*

To validate the theoretical predictions of the magnetic gating effect and first, to map the standing wave interaction, we experimentally measured the total luminescence of an ANP ($NaYF_4$: 5% $Tm^{3+}$/$NaYF_4$: 20% $Gd^{3+}$ core/shell) monolayer as a function of the mirror-sample distance (Z). Figure 3A displays these spatial scans for three distinct excitation powers (see additional details in the Supplemental Information). The luminescence exhibits clear, periodic oscillations corresponding to the electric field distribution intensity of the standing wave profile. The shaded background zones indicate the local field topology: blue zones correspond to electric-field antinodes (max of electric intensity), while green zones correspond to magnetic-field antinodes (max of magnetic intensity).

Consistent with our model, the emission maxima align strictly with the electric areas (blue), where the ESA cycling is most efficient. Conversely, the luminescence dips occur in the magnetic zones (green), where the purely electric ESA transition is inhibited. The contrast of these oscillations persists across all three power levels, confirming stable positioning within the interference pattern.

Figure 3B presents the central result of this study: the experimental quantification of the avalanching dynamics. By fixing the mirror position at a maximum of electric field (blue data) and a magnetic maximum (green data), we measured the characteristic power curves (Luminescence vs. Power, log-log scale) and compared them to the reference case without the mirror (red data).

The results show a pronounced divergence in behavior. In the reference free-space mode (red), the threshold shifts to higher powers and the nonlinearity moderates to s = 35. In the magnetic-gated regime (green), the avalanche dynamics behavior is strongly suppressed: the threshold is pushed to the highest power levels, and the nonlinearity is reduced by half, while in the purely electric regime (blue), the avalanche triggers at a low power threshold and rises with extreme steepness, reaching a 100% increase of the nonlinearity.

Surprisingly, we saw much greater modulation of slopes than expected between the experimental values and the theoretical predictions which can be attributed by to two key factors: (1) the GSA branching ratio (assumed as 75% electric / 25% magnetic) is highly sensitive to the local crystal field,[33, 34] which may deviate from idealized parameters; and (2) our analytical standing-wave model does not fully account for spatial averaging over the nanoparticle volume, which would require numerical simulations of the exact vector field distributions (see Supplementary Information). Nevertheless, the excellent qualitative agreement—such as the inverse correlation between threshold and s—confirms the validity of the magnetic-field gating mechanism.

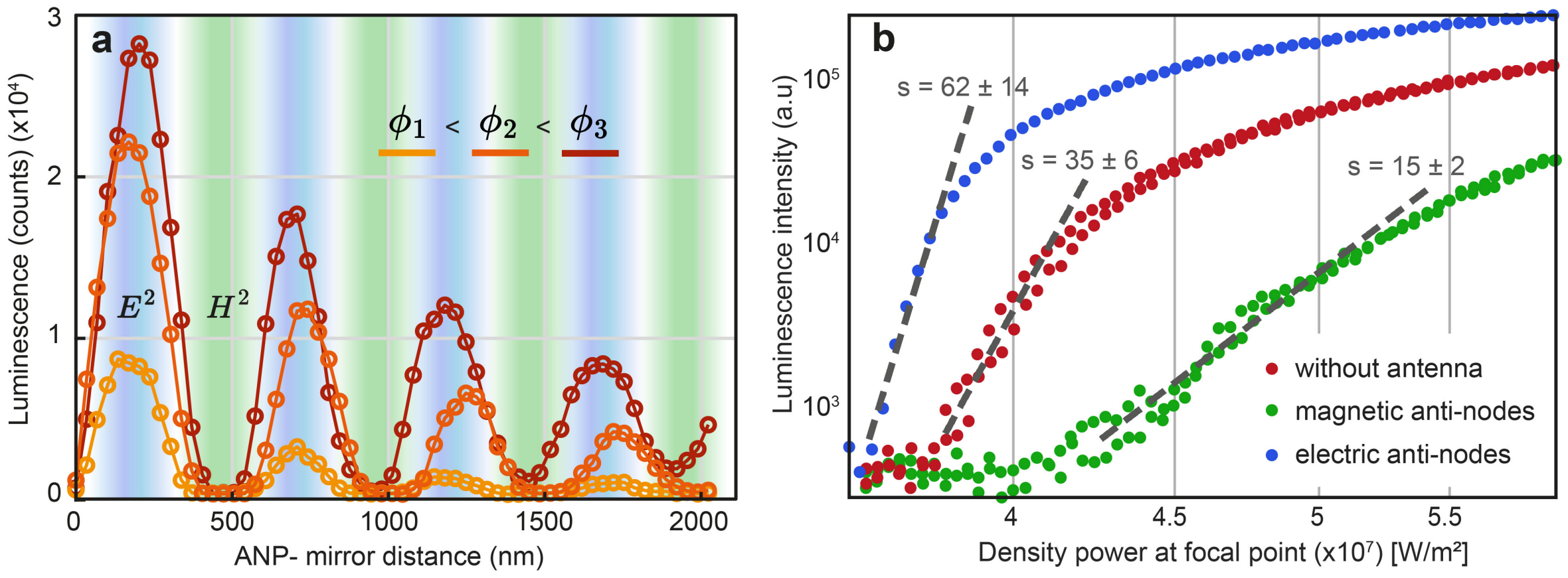

**Fig. 3.** ***Experimental demonstration of giant nonlinearity tuning.*** *a) Mapping the field interaction. Total luminescence of the nanoparticle monolayer as a function of mirror-sample distance (Z) for three excitation powers. The PA emission maxima align with electric-field antinodes (blue shaded zones), while minima correspond to magnetic-field antinodes (green shaded zones) where the ESA cycle component is inhibited. b) Quantification of gated dynamics. Experimental log-log scale s-curves*

*(Luminescence vs. Power) measured at fixed mirror positions. The system exhibits a giant dynamic range of control over both the avalanche threshold and the steepness of the transition: the electric-gated mode (blue) yields a low power threshold alongside a nonlinearity of s=62, the reference free-space mode (red) gives s=35, and the magnetic-gated mode (green) pushes the onset to higher intensities while suppressing the degree of nonlinearity down to s=15. The experimental data stands in excellent agreement with the theoretical predictions in Fig. 2.*

To demonstrate a concrete application of this electromagnetic-gated nonlinearity, we harness the spatial partitioning of electric and magnetic fields in the far field to push the limits of super-resolution optical imaging. Photon avalanching inherently enables sub-diffraction imaging due to the extreme nonlinearity of its optical response, which shrinks the effective point-spread function proportionally to the square root of the nonlinearity order ($1/\sqrt{s}$). In conventional scanning microscopy, however, this resolution remains bounded by the intrinsic nonlinearity of the material under standard illumination, where the electric and magnetic field components are spatially overlapped. By tailoring the vectorial topology of the excitation beam tightly focused through a high-numerical-aperture objective (NA = 1.30), we spatially disentangle the electric and magnetic field intensity distributions at the focal plane, thereby dynamically modulating the local degree of nonlinearity experienced by a single $Tm^{3+}$-doped nanocluster as it is raster-scanned across the focus.

Figure 4 illustrates this concept across three distinct illumination modes, whose theoretical focal intensity distributions are shown in Figures 4a–c (see Supplementary Information for details on the generation and modeling of the cylindrical vector beams). In the reference configuration (Figure 4a, left column), the nanocluster is illuminated by a fundamental Gaussian beam in which the electric and magnetic fields are spatially superimposed at equipartition. In the second configuration (Figure 4b, middle column), we employ a radially polarized cylindrical vector beam. Under tight focusing, this mode concentrates a pure longitudinal electric field component at the center of the focal spot, surrounded by a peripheral ring of magnetic field. As the nanocluster is scanned across this focal landscape, the local electric-to-magnetic field ratio—and consequently the avalanche nonlinearity order s—varies with position, reaching maximum nonlinearity right at the center of the beam where the electric field is isolated. Conversely, the third configuration (Figure 4c, right column) utilizes an azimuthally polarized cylindrical vector beam, which produces the dual electromagnetic topology: the magnetic field is confined at the focal center, while the electric field forms a surrounding ring.

Figures 4d–f present the experimental normalized luminescence maps obtained by scanning the exact same $NaYF_4$:5%$Tm^{3+}$/$NaYF_4$:20%$Gd^{3+}$ nanocluster across these three excitation profiles at an incident power set near the onset of the avalanche threshold, where the nonlinearity order s is highest. Under Gaussian illumination (Figure 4d), the intrinsic nonlinearity of the photon avalanche yields a sub-wavelength emission spot with characteristic full-width at half-maximum (FWHM) dimensions of about 264 nm and 375 nm along its two principal axes. Remarkably, under radial illumination (Figure 4e), the spatial isolation of the electric field at the focal center boosts the local avalanche nonlinearity, compressing the luminescence distribution down to about 180 nm by 204 nm. As detailed in the Supplementary Information, this substantial resolution enhancement far exceeds the mere geometric reduction of the longitudinal electric-field spot size relative to a Gaussian mode; it stems directly from the strong enhancement of the nonlinearity order s driven by the local suppression of the magnetic field at the beam center. In stark contrast, under azimuthal polarization (Figure 4f), the pure magnetic field at the center inhibits the electric-dipole-driven amplification loop (ESA + CR), suppressing central emission and producing a donut-shaped luminescence ring that maps the annular electric field distribution.

Increasing the excitation power toward the saturation regime (Figures 4g–i) progressively reduces the local nonlinearity slope along the response curve. While the emission profiles under Gaussian and radial illumination remain sub-wavelength due to the residual nonlinearity, their spatial footprints broaden (Figure 4g,h), respectively. Concurrently, in the azimuthal mode (Figure 4i), the luminescence ring thickens inward, shrinking the central dark region. Finally, driving the nanocluster into the saturation plateau further diminishes the effective nonlinearity, expanding the emission distributions for the Gaussian mode (Figure 4j), and for the radial mode (Figure 4k), as well as for the azimuthal mode (Figure 4l), where the central dark core is further reduced. Together, these results provide a concrete demonstration that spatially shaping the electric and magnetic field components to manipulate photon avalanche dynamics unlocks sub-wavelength imaging resolutions well beyond the conventional limits of photon avalanching.

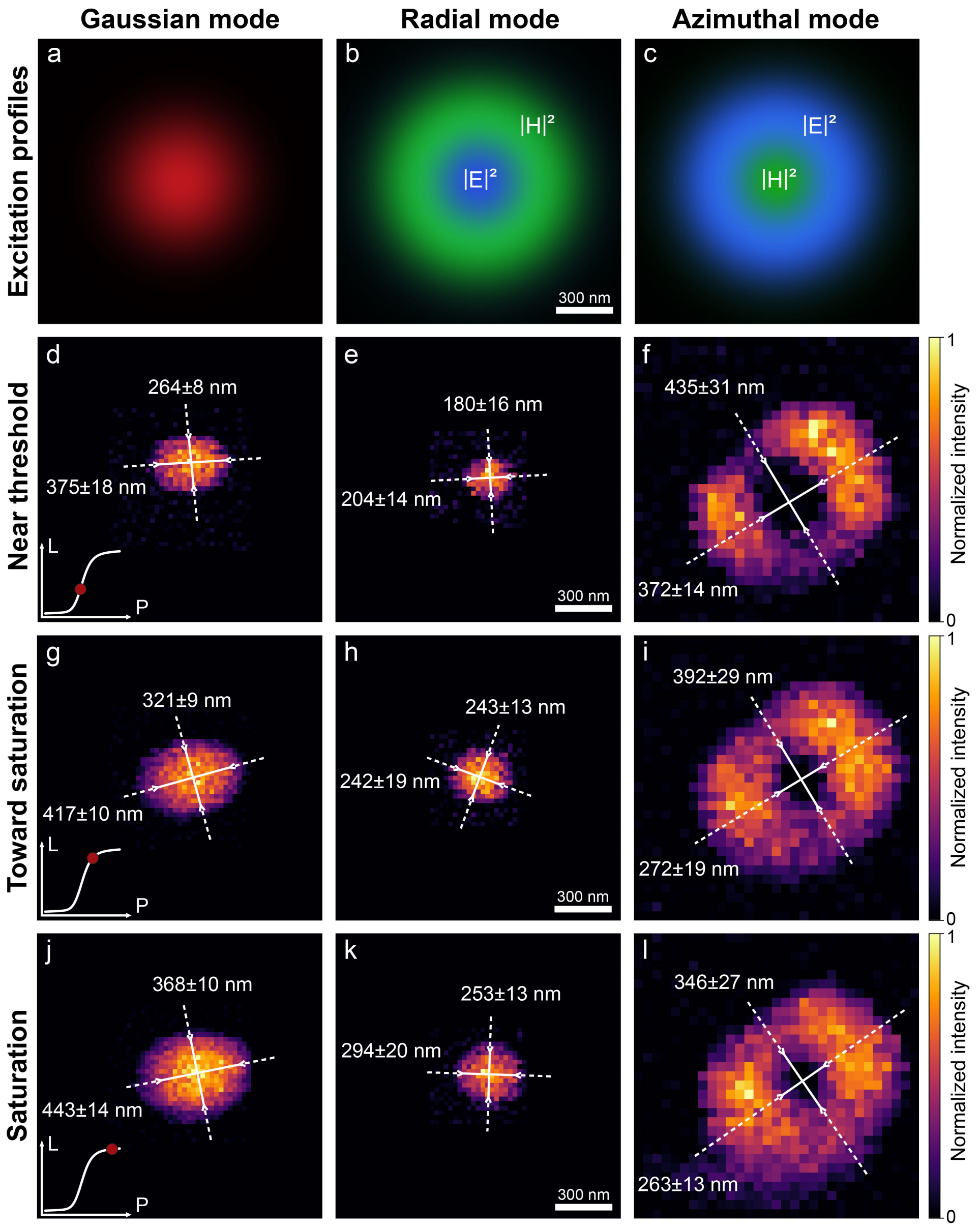

***Fig. 4. Enhanced super-resolution imaging via electromagnetic-field-gated photon avalanche nonlinearity.*** *a–c) Theoretical focal intensity distributions for three excitation modes: a) a fundamental Gaussian beam where electric and magnetic field components are spatially overlapped in equal proportion (red), b) a radially polarized cylindrical vector beam confining the electric field intensity (blue) at the center surrounded by a magnetic field intensity ring (green), and c) an azimuthally polarized cylindrical vector beam isolating the magnetic field at the center within an electric field ring. d–f) Experimental normalized luminescence maps obtained by scanning a single* $Tm^{3+}$*-doped nanocluster across the respective excitation profiles (a–c) at an excitation*

*power right at the onset of the avalanche threshold (Near threshold, where the nonlinearity order s is maximal), achieving a higher sub-wavelength resolution under radial polarization (e) compared to under Gaussian illumination (d), and an annular emission profile under azimuthal polarization (f). g–i) Corresponding luminescence maps acquired at an intermediate power approaching saturation, showing a progressive broadening of the emission spots as the local nonlinearity decreases. j–l) Luminescence maps recorded in the saturation regime, yielding characteristic showing a further broadening of the emission spots. The same nanocluster was used for all images (d–l). Insets in (d, g, j) schematically indicate the operating point (red dot) along the photon avalanche threshold curve (luminescence L versus excitation power P). All luminescence maps are normalized from 0 to 1, and all panels share the same spatial scale (scale bars: 300 nm).*

In conclusion, this work challenges the long-standing electro-centric dogma of optical interactions. By promoting the optical magnetic field from a neglected byproduct to a decisive control parameter, we have unlocked a new degree of freedom in quantum nanophotonics. The ability to actively control both the avalanche threshold and the steepness of the transition (sweeping the nonlinearity from $s = 15$ to $s = 62$)—simply by shaping the local electromagnetic landscape—demonstrates that avalanche dynamics can be a dynamically optically engineerable state. We directly translated this capability into a proof-of-concept super-resolution imaging scheme using cylindrical vector beams, demonstrating that spatially partitioning the electric and magnetic fields locally amplifies the avalanche nonlinearity of a single nanocluster to achieve sub-wavelength resolution (180 nm) markedly superior to conventional photon-avalanching nanoscopy in the same conditions. Beyond super-resolution imaging, this electromagnetic control of photon avalanche unlocks a broad spectrum of transformative applications in optical logic, sensing, and neuromorphic photonics. In neuromorphic computing, our magnetic-gating mechanism enables optical neurons with the dynamic plasticity essential for adaptive learning architectures. Furthermore, by biasing the system near its instability point, we open the door to "critical sensing" schemes, where the divergent susceptibility of the avalanche amplifies minute environmental perturbations. Finally, this platform establishes a bridge between solid-state physics and the universal dynamics of complex networks. Ultimately, our results suggest that the full potential of light-matter control lies not merely in the intensity of the electric field, but in the structured partition of its magnetic counterpart.

## 4. Materials and methods

### Nanomirror fabrication

The fabrication of the nanostructured probes followed a sequential process. First, an optical fiber was tapered using a Sutter P-2000 puller. To facilitate subsequent Focused Ion Beam (FIB) milling, a sacrificial 5 nm platinum layer was deposited. The tip was then precision-milled to create a planar end-face with a diameter of approximately 1 µm. Following this, an 80 nm-thick gold layer was thermally evaporated onto the processed tip, atop a 5 nm chromium adhesion layer. Finally, the roughness and diameter of the resulting nanomirror were verified using Scanning Electron Microscopy (SEM).

### Nanoparticles synthesis

ANPs of NaYF4: 5% Tm3+/NaYF4: 20% Gd3+ core/shell composition were prepared following previously established protocols with modifications.[40] Briefly, NaYF4: Tm3+ core nanocrystals were synthesized by heating lanthanide chloride, sodium oleate, and ammonium fluoride precursors in high-boiling-point solvents. After isolating and washing the core nanocrystals, they were overcoated with NaYF4:Gd3+ shells using a laboratory automation robot, WANDA. The final core/shell ANPs were washed and stored in hexane for further experiments. A detailed synthesis procedure is given in the Supplementary Information.

### Sample preparation

The doped nanoparticles are deposited by spin-coating on a glass substrate, previously cleaned by sonic bath and plasma cleaner. SEM observations allow to find the first and single layer of ANPs to conduct the experiments.

### Optical and near-field experimental setup

The $Tm^{3+}$-doped nanoparticles are excited using an infrared (IR) laser, which is filtered through a bandpass filter and a fixed polarizer. The incident power is controlled using a half-wave plate coupled with a polarization beam splitter cube (PBS). The angle of the half-wave plate is precisely adjusted via a digital interface. After the PBS, a fraction of the power is sampled using a simple coverslip and measured with a power meter. By fitting the data to Malus's law, a calibration is performed before measurements to establish the relationship between the half-wave plate angle and the incident power.

The optical near-field microscope (NT-MDT Integra) is mounted on an inverted microscope (Olympus IX73), with the nano-antenna-supported tip attached to a tuning fork vibrating at a frequency of 32 kHz. The tip's approach and feedback loop in the near field are controlled by monitoring the oscillation phase of the tuning fork, ensuring sub-nanometer oscillation amplitudes (below 1 nm). The tip is then aligned to the center of an oil-immersion objective (Olympus PLN 100x, NA 1.30), precisely over the study area excited using piezoelectric displacements in the XY plane of the SNOM head. The Z-axis displacement allows for positioning the antenna in the near field and maintaining nanometric control over the distance between the ANP layer and the nanomirror.

The luminescence is collected from below and can be analyzed in two different ways. It can be directed to a spectrometer (Sol Instruments MS5204i) after passing through a dichroic filter and a high-pass filter. The luminescence spectra are then measured using a CCD camera (Andor iDus 401 CCD) for each ANP layer-antenna position. Alternatively, the luminescence can be sent directly to an APD detector (micro photon devices PDM, PD-100-CTD) after passing through two successive filters (Thorlabs FESH1000 and Corion LP600) for higher sensitivity detection

**Availability of data and material**

The materials, data and code underlying the results presented in this paper are not publicly available at this time but may be obtained from the authors upon reasonable request.

**Supplementary Information**

Information for a detailed description of the experimental near-field and far-field optical configurations, including the generation of cylindrical vector beams for super-resolution imaging. The supporting material also presents the complete system of Differential Rate Equations (DRE) governing the five-level dynamics of the $Tm^{3+}$ ions, detailing the specific parameters for absorption, decay, and cooperative inter-ion mechanisms utilized in the theoretical model. Additionally, it provides the microscopic framework for spatially partitioned electromagnetic field excitation, alongside vectorial diffraction simulations used to explicitly disentangle pure geometric focusing effects from the fundamental electromagnetic enhancement of the avalanche nonlinearity.

**Acknowledgements**

This work is supported by the ERC grant FemtoMagnet (grant no. 101087709). We acknowledge the financial support from the Agence Nationale de la Recherche (ANR-20-CE09-0031-01, ANR-22-CE09-0027-04 and ANR-23-ERCC-0005), the Institut de Physique du CNRS (Tremplin@INP 2020). Work at the Molecular Foundry was supported by the Office of Science, Office of Basic Energy Sciences, of the U.S. Department of Energy under Contract No. DE-AC02-05CH11231. N.F.-M. acknowledges support from the European Union's Horizon 2020 research and innovation programme under the Marie Skłodowska-Curie grant agreement number 893439, the ISEF – de Gunzburg International Science Fellowship Program, and the Weizmann Institute's Women's Postdoctoral Career Development Award. A.S. acknowledges support from the European Union's Horizon 2020 research and innovation programme under the Marie Skłodowska-Curie grant agreement no. 895809 (MONOCLE).

**Author Contributions**

B.R and F.C conducted the experiments, developed the corrected theoretical model, and handled the simulations. F.C fabricated the nanomirror. E.C and A.S synthesized the $Tm^{3+}$ doped-nanoparticles. B.R, N.F.-M, S.B, P.J.S, and M.M design the experiment. The data analysis was performed by B.R, B.B, M.S.P, N.F.-M, S.B, J.P.S, and M.M. The article was written by B.R and M.M, incorporating contributions from all involved authors. N.F.-M, B.R and M.M. conceived the project.

**Conflict of interest**

The authors declare no conflicts of interest.